\documentstyle[preprint,revtex]{aps}

\begin{document}
\draft

\begin{title}
Exact diagonalization study of the two-dimensional $t-J-$Holstein \\
model
\end{title}

\author{A. Dobry, A. Greco, S. Koval and J. Riera}

\begin{instit}
Instituto de F\'{\i}sica Rosario, Consejo Nacional de Investigaciones
Cient\'{\i}ficas y T\'ecnicas \\
y Departamento de F\'{\i}sica, Universidad Nacional de Rosario, \\
Av. Pellegrini 250,  2000-Rosario, Argentina.
\end{instit}

\receipt{today}

\begin{abstract}
We study by exact diagonalization the two-dimensional
$t-J-$Holstein model near quarter filling by retaining only few
phonon modes in momentum space. This truncation allows us to
incorporate the full dynamics of the retained phonon modes.
The behavior of the kinetic energy, the charge structure factor and
other physical quantities, show the presence of a transition from a
delocalized phase to a localized phase at a finite value of the
electron-phonon coupling. We have also given some indications that
the $e$-ph coupling leads in general to a suppression of the pairing
susceptibility at quarter filling.
\end{abstract}


\newpage

The $t - J$ model is one of the simplest models proposed to
understand the physics of strongly correlated electronic
systems.\cite{emery}
In particular, the two-dimensional $t - J$ model has been extensively
studied in the context of high-T$_c$ superconductivity
since it contains the essential low-energy physics of the CuO$_2$
planes present in the cuprates. Many magnetic,
transport and superconducting properties of these materials have
been described by this model.\cite{dagorev} However, since this model
contains only electronic degrees of freedom, it cannot account for
other properties, like the isotope effect, which requires the
coupling of electron and phonon degrees of freedom.
The inclusion of phonon degrees of freedom could also be relevant to
describe electronic transport properties.
Moreover, given the basic nature of the $t - J$ model, it is of
general interest to study the effects of electron-phonon ($e$-ph)
coupling in the framework of strongly correlated systems.

Our starting point is the two-dimensional (2D) $t - J$ model. Exact
diagonalization studies of this model on the square lattice have
given strong unbiased indications of d$_{x^2-y^2}$ superconductivity
at quarter filling ($n = 0.5$) in the vicinity of phase
separation.\cite{dagrieprl,ohta} These indications are absent in the
one-band 2D Hubbard model, which is the most basic model of strongly
correlated electronic systems.

The $e$-ph coupling is incorporated in the Hamiltonian by a
Holstein term in which an Einstein phonon with
frequency $\omega$ is coupled linearly to the local electronic
density.  The Holstein model, which describes uncorrelated electrons
with this type of $e$-ph coupling, has been studied to
understand the relationship between superconductivity and Peierls
charge-density-wave (CDW) formation. Quantum Monte Carlo simulations
have been done in one dimension\cite{hirsch} and more recently in two
dimensions.\cite{scalettar,noack} In one dimension, it was shown in
Ref. \cite{hirsch} that the Holstein model undergoes a
transition to a CDW state above a critical value of the $e$-ph
coupling in the case of spinless fermions, but the ordered state
sets in for arbitrary value of the coupling for the case of
spin-$\frac{1}{2}$ fermions.

The total Hamiltonian including both electron and phonon degrees of
freedom is:
\begin{eqnarray}
\label{hamtjh}
H = - t \sum_{<i,j>, \sigma}
            ( \tilde{c}_{j \sigma}^\dagger \tilde{c}_{i \sigma}
            + \tilde{c}_{i \sigma}^\dagger \tilde{c}_{j \sigma} )
   + J \sum_{<i,j>}
           ( {\bf S}_{i} \cdot {\bf S}_{j}
           - \textstyle{1 \over 4} n_{i} n_{j} )   \\
\nonumber
   + \omega \sum_{i}
            ( b_{i}^\dagger b_{i} + \frac{1}{2})
   + g \sum_{i}
            ( b_{i}^\dagger + b_{i} )  n_i
\end{eqnarray}
\noindent
where $\tilde{c}_{i \sigma}^\dagger$ is an electron creation operator
at site $i$ with spin $\sigma$ with the constraint of no double
occupancy, $n_i = n_{i,\uparrow} + n_{i,\downarrow}$ is the electron
number operator, and $b_i^\dagger$ is a phonon creation operator at
site $i$. Here $t$ is the hopping
parameter, which we choose as the scale of energies, $J$ is the
antiferromagnetic exchange interaction and $g$ is the $e$-ph
coupling. We adopt $\omega = 1.0$ throughout.

The model defined by (\ref{hamtjh}) is usually called the
$t- J - {\rm Holstein}$ model. Both this model and the
Hubbard-Holstein model have been studied with a variety of analytical
and numerical methods.\cite{trapper,roder} Most exact diagonalization
studies have been performed so far for static phonons, in the
so-called adiabatic or frozen phonon approximation.\cite{roder}
The dynamics of phonons are therefore missed in these studies. Thus,
their results for many physical quantities that require the inclusion
of the full quantum nature of phonon degrees of freedom are
questionable.

The purpose of this report is to extend the exact diagonalization
techniques used for the pure $t - J$ model, in particular the
Lanczos method, to the $t- J - {\rm Holstein}$ model, preserving
the dynamics of phonons. This extension cannot be done without
further approximation because the bosonic part of the Hilbert space
is infinite dimensional. A truncation of the Hilbert space of
phonons in real space was proposed in Ref. \cite{fehrenbacher}.
In this approach, only two states are allowed for each mode.
Although the choice of these states in terms of coherent states is
pressumably optimal,\cite{zheng} it would be extremely hard to check
the validity of the resulting variational space. Alternatively,
we propose in this report a truncation in momentum space of phonons,
retaining a small number of phonon modes.
This approximation allows us to take enough boson states for each
of the retained phonon modes so that the physical quantities can
converge to the exact values for the most interesting region of
parameters.\cite{mustre}
The exact results thus obtained for a restricted set of phonon
modes in momentum space could be used to check approximate
methods like e.g. the one proposed in Ref. \cite{fehrenbacher}.

In order to present our approach, we rewrite the last two terms of
Eq. (\ref{hamtjh}) in momentum space as:
\begin{eqnarray}
\label{hamq}
H_{ph} =  \sum_{\bf q} \omega({\bf q})
            ( b_{\bf q}^\dagger b_{\bf q}  + \frac{1}{2})) \\
\nonumber
H_{e-ph} = \sum_{\bf q} g({\bf q}) ( b_{\bf q}^\dagger + b_{\bf -q} )
           \sum_{j} n_j e^{i {\bf q} . {\bf r}_j}.
\end{eqnarray}
\noindent
Actually, Eq. (\ref{hamq}) is a generalization of the respective
terms in (\ref{hamtjh}) because we have allowed a ${\bf q}$
dependence in the constants $\omega$ and $g$. In fact,
this ${\bf q}$ dependence
leads to a more realistic $e$-ph coupling for many materials. In the
case of the cuprates, starting from the three band Hubbard model in
the presence of a planar breathing-like phonon, an effective $e$-ph
coupling was derived with the form of Eq. (\ref{hamq}) where
$\omega({\bf q})$ and $g({\bf q})$ are maximum at
${\bf q} = (\pi,\pi)$.\cite{shepans} The different strength of the
$e$-ph coupling for different phonon modes in momentum space
provides a physical justification of our truncation scheme. That is,
we only keep in the phonon Hilbert space those states of the
phonons with largest $e$-ph coupling strength. In the
following, and for the sake of simplicity, we consider a
$g$ independent of the phonon momentum.

Let us consider the case in which only states of the phonon
${\bf q} = (\pi,\pi)$ are contained in the phonon part of the Hilbert
space.
The exact ground state of the Hamiltonian (\ref{hamtjh}) and
(\ref{hamq}) can be easily obtained for the case in which
$J = t = 0$, for an arbitrary electronic state \{$n_{i,\sigma}$\}.
This ground state is minimum when all electrons, if possible, are
located in one of the two sublattices of the square lattice. Then,
for finite $t$ and $J$,
a $(\pi,\pi)$ CDW order is expected when $g$ is large enough.
In Fig. \ref{fig1}, we show the ground state energy vs. number of
states of this phonon for the tilted $\sqrt{10} \times \sqrt{10}$
square lattice with $n = 0.6$, $J = 1.0$, and several
values of $g$. The convergence to the exact result is quite fast
in the range
$0.0 < g < 0.55$. Between $0.55 < g < 0.6$ the transition to the
above mentioned CDW region occurs. This convergence rate is typical
of many exchange couplings and fillings studied.

In the first place, we study various physical quantities in the
$\sqrt{10} \times \sqrt{10}$ cluster at $n = 0.6$ as a function
of $g$.
Their behaviour, shown in Fig. \ref{fig2}, indicate a transition at
$g=g_c$ from a phase which is essentially the same as in the pure
$t-J$ model ($g = 0$) to a radically different phase which is
essentially the one corresponding to $t = J =0$. The first phase
is usually called the ``delocalized" phase, and the second one, the
``localized" phase.
The kinetic energy is shown in Fig. \ref{fig2}(a).
It has a slow variation for $0 < g < g_c$. For
$g > g_c$, the kinetic energy decreases rapidly to zero indicating a
strong electron localization. We discuss further this feature
below. The transition to the localized region can be seen quite
clearly in the behavior of
the $(\pi,\pi)$ CDW structure factor defined as:
\begin{eqnarray}
\chi_{c} ({\bf q}) = \frac{1}{N} \sum_{i,j} <n_i n_j>
e^{i {\bf q . (r_i-r_j)}}
\label{suscdw}
\end{eqnarray}
\noindent
with ${\bf q} = (\pi,\pi)$, and in the occupation number of phonons
$n_{ph} = < b_{\bf q}^\dagger b_{\bf q} >$, shown in
Figs. \ref{fig2}(b)
and \ref{fig2}(c) respectively. The presence of two phases can also
be observed in other quantities like the spin structure factor at
$(\pi,\pi)$, shown in Fig. \ref{fig2}(d).

The transition to the CDW state is similar to that observed in
the spinless Holstein model by Hirsch and Fradkin.\cite{hirsch}
These authors derived an effective $\tilde{t} - V$ model, where $V$
is an effective nearest neighbor (nn) repulsion, to describe this
transition. $\tilde{t}$ is a renormalized hopping ($\tilde{t} \ll t$)
that contains the effect of increased effective mass of the charge
carriers. The fact that in our results the kinetic energy goes to
zero inmediately after the transition is because most of the
electrons are located in one of the two sublattices as it was
indicated above and this state is deeply buried energetically.
This feature should disappear if states of additional phonon modes
are included in the Hilbert space. In order to check this scenario
we have performed calculations for the same cluster and filling
by retaining three phonon modes in momentum space.
The three phonon modes are $(\pi,\pi)$, and
$(4\pi/5,2\pi/5)$ and $(-4\pi/5,-2\pi/5)$ which belong to the
closest shell to $(\pi,\pi)$. In this case, we obtained a
convergence of four digits in the kinetic energy with 4 states per
phonon mode for $g \le 0.3$, and 5 states per phonon mode for
$0.4 \le g \le 0.6$. The results for the kinetic energy for
various values of $J$ are shown in Fig. \ref{fig3}(a). We have
also reproduced for comparison the curves for $J = 1.0$
and $J = 3.0$ obtained by the phonon $(\pi,\pi)$ alone.
Since the phononic Hilbert space is enlarged with respect to the
previous one, the energy of the system in the region where phonons
dominate should be lowered, and a simple variational argument then
explains
why the transition occurs at a smaller $g_c$. As expected,
the kinetic energy decreases more slowly after the transition.
The transition is smoother and the influence of the $e$-ph
coupling is noticeable at much smaller $g$. For fixed
$J$ and $g$, the kinetic energy is smaller than the corresponding
value obtained for the case of a single mode, indicating a larger
increase of the effective mass of the carriers. In Fig. \ref{fig3}(b)
it can be seen that the charge structure factor at
${\bf q} = (3 \pi /5, -\pi / 5)$, which is the largest at $g =0$
in the
binding region ($J \ge 2.0$), is suppressed by $g$ and the structure
factor at ${\bf q} = (4 \pi /5, 2 \pi / 5)$ is enhanced and
becomes dominant for $g > g_c$. Thus, the
localized region is an incommensurate CDW order.

As in Ref. \cite{hirsch}, it is reasonable to assume that our
electron-phonon model can be described by an effective, purely
electronic, $t-J-V$ model. In this model, the effective nn
repulsion $V$ acts against binding of holes on nearest
neighbor sites, thus suppressing pairing but then also preventing
phase separation (PS). The outcome of these competing effects is not
easily predictable and numerical calculations have in fact indicated
an enhancement of pairing correlations.\cite{dagrietjv}
In the $t-J-$Holstein model, we have an additional competing effect,
i.e. the
renormalization of the mass of the carriers. In order to determine
the result of these effects, we have computed the kinetic energy
and the pairing susceptibility for the $4 \times 4$ lattice at
quarter filling.
The static pairing susceptibility is defined as:
\begin{eqnarray}
\chi_{SC} = \frac{1}{N} \sum_{i,j} <\Delta_i^\dagger \Delta_j>
\label{suspair}
\end{eqnarray}
\noindent
where
\begin{eqnarray}
\Delta_i = \sum_{\mu} g_\mu c_{i\uparrow} c_{i+\mu\downarrow}
\end{eqnarray}
\noindent
where the sum extends over the nearest neighbors of site $i$ and
$g_\mu$ are the form factors that determine the pairing symmetry.
We report here results for the extended-s and d$_{x^2-y^2}$ pairing
symmetries.

We have considered only the phonon $(\pi,\pi)$ in the calculation.
The exact diagonalization of the $4 \times 4$ cluster at quarter
filling, was performed with momentum (0,0) and d-wave rotational
symmetry. The spin reversal symmetry was also taken into account
to reduce the
dimension of the electronic part of the Hilbert space. The energy
converges with approximately 20 states for the phonon mode.
Let us first examine the results for the kinetic energy,
shown in Fig. \ref{fig4}(a).
The general behavior is similar to that found for the
$\sqrt{10} \times \sqrt{10}$ lattice. However, a slight upward
change of the kinetic energy for $J = 3.5$ and $J=4$, for large $g$
can be seen. In the pure $t-J$ model, $J=3.5$ is precisely
at the PS border.\cite{dagrieprl}
The influence of the effective repulsive interaction is then to
shift the PS border to larger $J$.

Finally, in Fig. \ref{fig4}(b) we show results for the pairing
susceptibility. For the pure $t-J$ model, as shown in
Ref. \cite{dagrieprl}, the
d$_{x^2-y^2}$ pairing dominates over the extended-s pairing
and its maximum occurs for $J=3$, i.e. just before the PS border.
The effect of $e$-ph coupling is in general to suppress pairing.
Only for $J = 3.5$ and $J=4$ and large $g$, consistently with the
behavior of the kinetic energy, an enhancement of the pairing
susceptibility can be observed. If more phonon modes are added,
taking into account the results previously discussed, it is
expected that the effective nn repulsion turns out to be weaker
than for the one phonon case. Besides, the mass
renormalization is expected to be stronger. Thus, the net effect
would be an overall suppression of the pairing correlations and
susceptibility.

In summary, we have studied using exact diagonalization the 2D
$t-J-$Holstein model by retaining only few phonon modes in momentum
space. Near quarter filling, the behavior of the kinetic energy,
the charge structure factor, and other physical quantities, show
the presence of a transition from a delocalized phase to a
localized phase at a finite $g_c$. This value of $g_c$ decreases
when the number of phonon modes is increased. We have also given some
indications that the $e$-ph coupling leads in general
to a suppression of the pairing susceptibility at quarter filling.
Although preliminary, these are among the first results for a 2D
electron-phonon model which incorporates the full dynamics of the
phonon field and could be relevant to physically realistic situations.

\newpage

\figure{The ground state energy E$_0$ as a function of the
number of states of the $(\pi,\pi)$ phonon for the
$\sqrt{10} \times \sqrt{10}$ cluster with $n = 0.6$,
$J = t = \omega =1.0$, and several values of $g$.
\label{fig1}}

\figure{(a) Kinetic energy, (b) charge structure factor
$\chi_c(\pi,\pi)$, (c) phonon occupancy, and (d) spin structure
factor $\chi_s(\pi,\pi)$, as a function of the $e$-ph coupling,
obtained by exact diagonalization for the
$\sqrt{10} \times \sqrt{10}$ cluster with $n = 0.6$,
$t = \omega =1.0$, and several values of $J$. The
symbols used in (b), (c) and (d) are the same as in (a).
\label{fig2}}

\figure{(a)The kinetic energy as a function of $g$ for the
$\sqrt{10} \times \sqrt{10}$ cluster with $n = 0.6$,
$t = \omega =1.0$, and several values of $J$. Full lines
correspond to the set of $(\pi,\pi)$, $(4\pi/5,2\pi/5)$ and
$(-4\pi/5,-2\pi/5)$ phonons. Results for the $(\pi,\pi)$ phonon
alone (dashed lines) are included for comparison.
(b)The CDW structure factor at $(3\pi/5,-\pi/5)$ (full lines)
and at $(4\pi/5,2\pi/5)$ (dashed lines) as a function of $g$.
The coupling constants are the same as in (a).
\label{fig3}}

\figure{(a) Kinetic energy, and (b) pairing susceptibility for
$d_{x^2-y^2}$ (solid symbols) and extended-s (open symbols)
symmetries, for the $4 \times 4$ lattice at quarter filling and
$J=2.0$ (up triangles), $J=2.5$ (circles), $J=3.0$ (squares),
$J=3.5$ (diamonds) and $J=4.0$ (down triangles).
\label{fig4}}

\end{document}